# Reality without Realism: On the Ontological and Epistemological Architecture of Quantum Mechanics

## Arkady Plotnitsky[a] and Andrei Khrennikov[b]

[a] *Theory and Cultural Studies Program,*
*Purdue University, West Lafayette, IN 47907, USA*

[b] *International Center for Mathematical Modeling*
*in Physics, Engineering, Economics, and Cognitive Science*
*Linnaeus University Växjö-Kalmar, Sweden*

**Abstract**. First, this article considers the nature of quantum reality (the reality responsible for quantum phenomena) and the concept of realism (our ability to represent this reality) in quantum theory, in conjunction with the roles of locality, causality, and probability and statistics there. Second, it offers two interpretations of quantum mechanics, developed by the authors of this article, the second of which is also a different (from quantum mechanics) theory of quantum phenomena. Both of these interpretations are statistical. The first interpretation, by A. Plotnitsky, "the statistical Copenhagen interpretation," is nonrealist, insofar as the description or even conception of the nature of quantum objects and processes is precluded. The second, by A. Khrennikov, is ultimately realist, because it assumes that the quantum-mechanical level of reality is underlain by a deeper level of reality, described, in a realist fashion, by a model, based in the pre-quantum classical statistical field theory (PCSFT), the predictions of which reproduce those of quantum mechanics. Moreover, because the continuous fields considered in this model are transformed into discrete clicks of detectors, experimental outcomes in this model depend on the context of measurement in accordance with N. Bohr's interpretation and the statistical Copenhagen interpretation, which coincides with N. Bohr's interpretation in this regard.

## 1. Introduction

This article proceeds along two lines of inquiry. First, it considers the nature of quantum *reality* (the nature of objects and processes that are responsible for observed quantum phenomena) and various forms of *realism* (ways of representing this reality) in quantum theory.[1] Second, it offers

---

[1] The concepts of reality and realism will be explained in Section 2. By "quantum phenomena" we refer to those observable physical phenomena in considering which Planck's constant, $h$, must be taken into account, although, as will be seen below, these phenomena themselves may be described by means of classical physics. Quantum objects, which are responsible for the appearance of quantum phenomena, could be macroscopic, although their quantum nature would be defined by their ultimate microscopic constitution and hence by the role of $h$ in the corresponding phenomena. (We discuss the difference between objects and phenomena in quantum physics below). By "quantum theory" we refer collectively to theories accounting for

two interpretations of quantum mechanics, developed by the two authors of this article. (The second of these interpretations also arises from a different overall theory of quantum phenomena.)

The main questions at stake in the first, more general, line of our argument are a) whether it is possible to represent the ultimate nature of reality responsible for quantum phenomena, and b) whether quantum mechanics, in its standard version, is a realist theory. These questions came into the foreground early in the history of quantum mechanics, especially in the famous confrontation between A. Einstein and N. Bohr, which has shaped the subsequent debate concerning quantum mechanics and its interpretation. This debate still continues with undiminished intensity. Einstein argued that quantum mechanics is incomplete because it does not describe, at least not completely, individual quantum objects and processes, analogously to the way classical mechanics or relativity does, at least ideally and in principle, for the objects and processes it considers. We will term this concept of completeness "the Einstein completeness." Bohr counter-argued that, while Einstein's claim may be true, and while it is strictly true in Bohr's and related interpretations, in, as W. Heisenberg called it, "the Spirit of Copenhagen" [1, p. iv], such as the statistical Copenhagen interpretation proposed in Section 4 of this article, quantum mechanics may be seen as complete in a different sense. It is as complete as nature allows a theory of quantum phenomena to be within the proper scope of quantum mechanics, *at least as things stand now* (a crucial qualification assumed by Bohr and throughout this article), insofar as quantum mechanics correctly predicts the outcomes of all quantum experiments performed thus far. We term this concept of completeness "the Bohr completeness."

It is true, and was part of Einstein's concern, that, unlike those of classical mechanics and classical electrodynamics, or of relativity, where it is possible to predict the outcomes of individual physical processes (ideally) exactly, in quantum mechanics these predictions are, in general, probabilistic or statistical even in dealing with elemental individual quantum processes and events. This, however, is strictly in accord with what is actually observed, because identically prepared quantum experiments, in general lead to different outcomes [e.g., 2, v. 2, p. 73]. There is no kind of quantum event concerning which even ideally exact predictions are rigorously possible. It may be a matter of interpretation whether one could assign probabilities to the outcomes of individual quantum experiments or could only deal with the statistics of (multiple) repeated experiments, an alternative considered below. However, the generally probabilistic nature of quantum predictions is not in question, again, as things stand now, even if one offers an Einstein-complete theory of quantum objects and processes, such as one or another form of Bohmian mechanics, or of some deeper underlying dynamics, such as A. Khrennikov's model proposed here. The question, then, becomes whether nature will at some point allow us to have an Einstein-complete theory of quantum phenomena. Einstein thought it should. Bohr

---

quantum phenomena, among them the "standard quantum mechanics" (introduced by W. Heisenberg and E. Schrödinger in 1925-1926), henceforth designated as "quantum mechanics," in contradistinction, for example, to "Bohmian mechanics," which is a mathematically different theory, rather than a different interpretation of quantum mechanics. By "quantum physics" we refer to the totality of quantum phenomena and quantum theories. These concepts technically include high-energy (relativistic) quantum theories and physics, which will not be considered here. The terms "classical phenomena," "classical mechanics," "classical theory," and "classical physics," will be used in parallel. We are only concerned, in considering locality, with special relativity, and henceforth "relativity" refers to special relativity, unless stated otherwise.

thought it *might not*, which is not the same as it never will but which, Bohr contended, would at least diminish the force of Einstein's criticism [e.g., 2, v. 2, p. 57].

At a certain stage of this debate, in the 1930s, the question of locality was injected into it, due to several thought experiments proposed by Einstein, especially those of the EPR (Einstein-Podolsky-Rosen) type, first presented in the famous paper of A. Einstein, B. Podolsky, and N. Rosen (EPR) [3].[2] By using these experiments, Einstein argued that quantum mechanics could be considered as offering a complete description of the physical reality of individual quantum processes only if quantum mechanics or nature itself is nonlocal. Bohr counter-argued that, even given these experiments, quantum mechanics could be shown to be both complete, by his criterion (Bohr-complete), and local [4].[3] As did Einstein, Bohr ruled out the nonlocality of nature, again, at least on the basis of the evidence available thus far, which is one of the reasons why neither of them saw Bohmian mechanics (introduced in 1952) as a viable alternative. Bohr's argument, thus, left open the question whether a more complete and specifically an Einstein-complete local theory is possible, a question that remains open. Eventually, especially in the wake of the Bell and Kochen-Specker theorems, the question of locality, rather than of completeness, came to dominate the debate concerning quantum phenomena and quantum theory. However, because realism has remained a major concern, in particular given the lack of realism as a possible alternative to nonlocality, the question of completeness and the suitable criteria of completeness have remained germane to this debate as well. Indeed these theorems sharpened and gave more rigorous sense to the question of the completeness of quantum mechanics, and (in our terms here) the difference between Einstein completeness and Bohr completeness in considering it.

In sum, the quantum-mechanical situation may be seen as essentially shaped by three interrelated questions, each defined by a cluster of main concepts—(1) reality, realism, and completeness; (2) randomness and causality, and hence probability and statistics; and (3) locality. The corresponding assumptions and alternatives, such as realism vs. locality, define different interpretations of quantum phenomena and quantum mechanics, or of the Bell and Kochen-Specker theorems and related findings pertinent to quantum foundations.[4]

---

[2] We explain "locality" in more detail in the next section. For the moment, at stake is primarily the compatibility with (special) relativity, prohibiting an "action at a distance" (a physical influence propagating faster than the speed of light in a vacuum).

[3] For a detailed analysis of the exchange see [5, pp. 237-312].

[4] It is not possible to survey these interpretations here. Just as does the Copenhagen interpretation (one could think of several even in Bohr's own case), each rubric, on by now a long list (e.g.. the many-worlds, consistent-histories, modal, relational, transcendental-pragmatist, and so forth), contains numerous versions, which, however, revolve around the concepts just listed. The literature dealing with the subject is immense, although standard reference sources, such as *Wikipedia* [6], would list the most prominent ones. Given the role of the statistical considerations in this article, we might mention a compelling statistical interpretation proposed in [7], which has certain affinities with the statistical Copenhagen interpretation proposed here, again, acknowledging that there is a large number of other statistical interpretations available, difficult to survey here. The literature is nearly as immense when it comes to the Bell and the Kochen-Specker theorems and related findings. We will not be concerned with these problematics here. Among the standard treatments are [8, 9, 10]. We might add that they have been extensively discussed at Växjö conferences, and the proceedings of these conferences contain important

Both the interpretations of quantum mechanics to be offered in this article are statistical. The first interpretation, by A. Plotnitsky, "the statistical Copenhagen interpretation," is statistical because in this interpretation quantum mechanics is seen as a theory that only statistically predicts the data observed in *repeated* (identically prepared) experiments, rather than the probabilities of single experiments, which probabilities cannot, in general, be assigned in this interpretation. The second interpretation, by A. Khrennikov, is statistical, first, in the same sense as the statistical Copenhagen interpretation and, second, insofar as it is based in the "pre-quantum classical statistical field theory" (PCSFT). Thus, at stake in Khrennikov's overall mathematical model is also a different theory of quantum phenomena, rather than only a different interpretation of quantum mechanics, underlined by this theory.

These two interpretations are different as concerns their respectively nonrealist and realist nature. The first is strictly nonrealist because in this interpretation a description or, more radically (Bohr does not appear to claim as much, at least expressly), even a conception of the ultimate nature of quantum objects and processes is, in Bohr's words, "*in principle* excluded" [2, v. 2, p. 62]. It is also not entirely clear whether Bohr had subscribed to the view, adopted by the statistical Copenhagen interpretation, that one cannot even meaningfully assign probabilities to the outcomes of individual quantum processes and events. Given, however, that Bohr uniformly spoke of the *statistical* rather than *probabilistic* character of quantum predictions, and that he is very careful in choosing his terms, he might well have held this view. As will be seen, among the founding figures of quantum mechanics, W. Pauli appears to have expressly adopted this view.

The second interpretation or, again, the model that enables it is realist because it assumes that the level of reality corresponding to that considered by quantum mechanics is underlain by a deeper level of reality, described by a classical-like field theory or model (PCSFT). This model reproduces quantum probabilities, because the continuous fields considered are transformed into discrete clicks of detectors. As a result, experimental outcomes in this model depend on the context of measurement just as they do in Bohr's interpretation or in the statistical Copenhagen interpretation. It follows that, insofar as it is possible to speak of objectivity in either of these two interpretations, this objectivity is different from that found in classical physics or relativity, where objectivity is defined by the theory's capacity to relate to the data pertaining to the corresponding objects and processes themselves independently of the context of measurement.

The remainder of this article proceeds as follow. Section 2 offers an outline of key concepts considered here. Section 3 revisits the Bohr-Einstein debate and presents Bohr's interpretation of quantum phenomena and quantum mechanics, in Bohr's *ultimate* version, which confirms that there is no single Copenhagen interpretation even in Bohr's own case. As indicated above, however, it is possible to speak, with Heisenberg, of "the Spirit of Copenhagen" [1, p. iv]. Section 4 considers the statistical Copenhagen interpretation, by Plotnitsky, and Section 5 Khrennikov's PCSFT model and the interpretation of quantum mechanics based in it.

## 2. An Outline of Concepts

This section outlines the key concepts used by this article, *as these concepts will be understood here*, because they could be understood otherwise. Although this outline cannot claim to capture

---

contributions to the subject, and most other foundational issues concerning quantum physics. It is worth keeping in mind that these theorems and most of these findings pertain to quantum data as such, and do not depend on quantum mechanics or any particular theory of these data.

all of the deeper physical and philosophical aspects of these concepts, it is sufficient for our purposes and, hopefully, for avoiding misunderstandings concerning these concepts and our use of them. We begin with the concept of concept itself, a term often used without further explanation in physics or even philosophy. As understood in this article, in part following Deleuze and Guattari [11], a concept is not a generalization from particulars (which commonly defines concepts) or merely a general or abstract idea, although a concept may contain such generalizations and abstract, including mathematical, ideas. A concept is a multi-component entity, defined by the specific *organization* of its components, which may be general or particular, and some of these components, indeed often the most important ones, are concepts in turn. It is the relational and hierarchical organization of these components that is most crucial in defining a concept. In practice, there is always a cut off in delineating a concept, which results from assuming some of the components of this concept to be primitive entities whose structure is not specified. These primitive concepts could, however, be specified by an alternative delineation, which would lead to a new overall concept, containing a new set of primitive, (provisionally) unspecified components. The history of a concept is often that of such progressive new delineations.

The concept of a moving body in classical physics is an example of a multi-component physical concept. It involves multiple (idealized) elements related to its properties and behavior, beginning with the concept of motion, defined by such component concepts as position and velocity or momentum, mathematized by means of differential functions of real variables. This concept has its history as well, beginning with Aristotle's concept of physical motion, lacking in mathematical architecture as it was. This architecture was introduced into the concept of motion by the modern classical physics of Galileo and Newton, as a mathematical-experimental science of nature, which changed this concept. This concept is very different from Bohr's concept of a quantum object, the properties of which cannot be specified in the way they are in classical physics. This impossibility gives Bohr's concept a special status insofar as this concept or, by extension, the quantum-level *reality* as understood here, has no conceptual architecture that is or could ever be associated with it. In Bohr's interpretation, defined by this concept, the mathematical concepts comprising the formalism of quantum mechanics only relate, in terms of probabilistic or statistical predictions, to what is observed in measuring instruments, which or, more accurately, their observable parts, are described by means of classical physical concepts, thus allowing for realism at this level. As explained below, measuring instruments also have quantum (and hence, unobservable) parts through which they interact with quantum objects.

We shall now define the concepts of reality and realism, actually two sets of concepts corresponding to these terms. These two sets of concepts are, we believe, sufficiently general to encompass most concepts of reality and realism currently used in physics and the philosophy of physics, and we will indicate some among more specific versions of both concepts, thus defined, as we proceed. By "reality" we refer to that which actually exists or is assumed to exist. In the case of physics, it is nature or matter, which is generally, but not always, assumed to exist independently of our interaction with it, and to have existed when we did not exist and to continue to exist when we will no longer exist. This is also true in Bohr's and related interpretations in the spirit of Copenhagen, such as the statistical Copenhagen interpretation, but in these cases in the absence of any description or even conception of the character of this existence, and hence of realism. What exists or is assumed to exist and how it exists is a matter of perspective, interpretation, and debate.

*Claims concerning* what exists and how it exists define the corresponding concepts of realism. It follows that, according to this definition, any form of realism is more than only a claim concerning the existence of something, such as physical objects, which we can describe, at least by means of idealized models, in classical physics, or quantum objects, about which nothing else could be said or even thought in Bohr's or related interpretations. By realist theories we understand theories of the following two types. According to *the first type of realism*, a realist theory would offer a representation of the properties of the physical objects or systems of objects considered and their behavior, or sometimes, as in the so-called structural realism, of the structures defining such systems. In modern (post-Galilean) physics, generally, such a representation is given in or by the mathematical formalism of the theory. This representation may be and in modern physics usually is an idealized and specifically mathematized representation or model of the actual reality, which retains some of the features of the actual objects and processes considered and disregard others. Realist theories are sometimes called ontological theories, although the term "ontological" may carry additional philosophical connotations, with which we shall not be concerned here.[5] One could, in principle, see claims concerning the existence of something to which a theory relate in any way, even without representing that something, as realism. However, placing this view of reality outside realism is consistent with a more common use of the term realism in physics and philosophy, and it is advantageous in the context of Bohr's and related interpretations, defined by this concept of reality as "reality without realism."

According to *the second type of realism*, one would presuppose an *independent architecture* (which may be temporal) of reality governing this behavior, even if this architecture cannot be represented, idealized, by a theory either at a given point in history or perhaps ever, but if so, only due to practical limitations. In the first of these two eventualities, a theory that is merely predictive may be accepted for the lack of a realist alternative, but under the assumption that a future theory will do better, specifically by virtue of being a realist theory of the first type, just defined. Einstein held this view toward quantum mechanics, which he expected to be eventually replaced by such a realist theory, ideally, a field theory of a classical type, on the model of Maxwell's electrodynamics, followed by Einstein's general relativity.

What unites both conceptions of realism and thus defines realism most generally is the assumption that this type of architecture exists independently of our interactions with it. In other words, realism is defined by an assumption, which defined Kant's philosophy, that the ultimate constitution of nature possesses attributes that may be unknown or even unknowable, but that are *thinkable*, conceivable [14, p. 115]. In physics, this constitution is often deemed conceivable on the model of classical physics and its ideal of reality, possibly adjusted to accommodate new phenomena, such as electromagnetic or relativistic ones, and new concepts, such as field, classical or quantum, or more recently automata, including quantum automata [15,16]. This architecture is, more commonly than not, assumed to be approachable, along the lines of the first type of realism defined above, again, on the model of classical physics or relativity. At the very least, realism assumes that the concept of organization can in principle apply to this constitution, no matter how much off the mark anything we can specifically come up with at a given point in conceiving of this constitution may be. The hope is, however, that our theories can eventually capture something of this architecture in one way or another, to one degree or another, at one distance from reality or another.

---

[5] The description just given allows for different degrees to which our models "match" reality. For example, to what degree does the mathematical architecture of relativity correspond, even as an idealization, to the architecture of nature, as opposed to ultimately only serving as a mathematical model for correct predictions concerning relativistic phenomena? See [12]. Indeed, as Kant realized, these questions could be posed concerning classical mechanics, where, however, the descriptive idealizations used are more in accord with our phenomenal experience than in relativity or quantum theory. There is vast literature on the subject, which we cannot consider here. One might mention, however, E. Schrödinger's account, arguably following H. Hertz, of "the physics of models" in classical physics in his cat-paradox paper, which addressed the limitations of our capacity to represent, to have a picture, *Bild*, of the ultimate reality even if the latter is assumed to be classical, which Schrödinger preferred it to be [13, pp. 152-153]. While he was thus close to Einstein in preferring the "classical ideal" to what transpired in quantum mechanics, he was more skeptical than Einstein as concerns the future of this ideal in fundamental physics.

Interpretations of quantum phenomena and quantum mechanics in the spirit of Copenhagen, such as that of Bohr or the statistical Copenhagen interpretation proposed here, not only do not make any of these assumptions concerning quantum objects or their behavior, but also disallow all of these assumptions. These interpretations, again, assume that quantum objects, or some entities in nature that we thus idealize as quantum objects, do exist independently of our interaction with them, are *real*, and that it is this existence or reality that is responsible, through this interaction, for the situation we encounter in quantum physics. In other words, the character of this existence is such that it preludes us from describing quantum objects and their behavior, or even from forming a conception of them. As indicated above, Bohr does not appear to have made or expressly stated the last strong claim, but this claim, which is part of the statistical Copenhagen interpretation, may be argued to be in the spirit of Copenhagen. The nonrealist argumentation just outlined (now including that of Bohr) is not simply "an arbitrary renunciation" of an analysis the ultimate constitution of nature [2, v. 2, p. 62]. It is an *argument* that, on the basis of its analysis of the nature of quantum phenomena and quantum theory, specifically quantum mechanics, is compelled to conclude that this constitution is beyond the reach of theoretical description or even thought itself, at least as things stand now. In this sense, one can speak of "reality without realism": quantum objects may said to be real and, as such, have effects on the world we observe, and yet prevent us from representing them, specifically by the formalism of quantum theory, or possibly even from forming any conception concerning them.[6]

_______________

[6] As noted earlier, once one allows for "reality" in this sense, one could, in principle, speak of "realism." I. Hacking's influential concept of "entity-realism" (vs. "theory realism") [17] and some of its avatars, developed during the last decade in the philosophy of science and debates concerning the question of reality there (debates influenced by quantum theory), may be argued to be examples of this type of realism. These concepts do not appear to us to be quite as radical as the concept of "reality without realism," insofar as they still appear to conform to Kant's concept of noumena as, while unknowable, still in principle thinkable. As noted here, Bohr's position could be interpreted in either direction. The subject would require a further discussion, which is beyond our scope here. It may be added that one could, following G. Berkeley, conversely, understand nonrealism as the denial of the existence of external reality altogether. While this concept, used by Berkeley against Newton, is not without relevance to quantum theory, it is rarely adopted in physics or philosophy, and will be not considered here.

There is still an ontology or realism associated with such interpretations. It is defined by the classical ontology of measuring instruments, which—that is, *their observable parts*—are described by classical physics and in which the outcomes of quantum experiments are registered as the effects of the interaction between quantum objects and these instruments. This interaction itself is quantum, and hence, it is not available to a realist treatment, but in each case, this interaction leaves, as its effect, a mark in a measuring instrument. This mark can then be treated as a part of a permanent record, which can be unambiguously defined, discussed, communicated, and so forth, and in this sense may be seen as objective. It is the specific character of these effects that compels nonrealist interpretations of quantum objects as entities that are beyond all description or even conception, as opposed to merely postulating this nature of quantum objects or what this concept idealizes.

The lack of causality is an automatic consequence. As Schrödinger observed, by way of a very different assessment of this type of argumentation, which he saw as "a doctrine born of distress," in his cat-paradox paper: "if a classical state does not exist at any moment, it can hardly change causally" [13, p.154]. We need, however, to define the concept of causality to which this assessment refers and that will be assumed here. In this view, "causality" is an ontological category (part of reality). It relates to the behavior of physical systems whose evolution is defined by the fact that the state of a given system is, at least at the level of idealized models, determined at all moments of time by their state at a particular moment of time, indeed at any given moment of time.[7] By contrast, "determinism" is assumed here to be an epistemological category (part of our knowledge of reality) that denotes our ability to predict the state of a system, at least, again, as defined by an idealized model, exactly, rather than probabilistically, at any and all points once we know its state at a given point. Determinism is sometimes used in the same sense as causality, as defined here, and in the case of classical mechanics (which deals with single objects or a small number of objects), causality and determinism, as defined here, coincide. Once a system is large enough, one needs a superhuman power to predict its behavior exactly, as was famously noted by P. S. Laplace.

However, while it follows automatically that noncausal behavior, *considered at the level of a given model*, cannot be handled deterministically, the reverse is not true. The underlined qualification is necessary because we can have causal models of processes in nature that may not be ultimately causal. Thus, the fact that the causal models of classical physics apply and are effective within the proper limits of classical physics does not mean that the ultimate character of the actual processes that are responsible for classical phenomena is causal. They may not be, for example, by virtue of their ultimately quantum nature. Nor, conversely, does the noncausal character of a model, for example, that of quantum mechanics in the spirit of Copenhagen, guarantee that quantum behavior is noncausal. It may ultimately prove to be causal.

---

[7] Sometimes the term "causality" is used in accordance with the requirements of relativity, which further restricts causes to those occurring in the backward (past) light cone of the event that is seen as an effect of this cause, while no event can be a cause of any event outside the forward (future) light cone of that event. These restrictions follow from the assumption that causal influences cannot travel faster than the speed of light in a vacuum, *c*. When speaking of the lack of causality, we only mean the inapplicability of the concept of causality found in classical physics and not about any incompatibility with relativity.

Noncausal models of the quantum constitution of nature are idealizations as well. Rigorously, it is only determinism that quantum phenomena preclude, because, as noted earlier, it is a well established experimental fact that identically prepared quantum experiments in general lead to different outcomes [2, v. 2, p. 73]. However, as against classical physics, the difference between these outcomes cannot be improved beyond certain limits (defined by the uncertainty relations), for example, by improving the precision of our instruments. This makes individual experiments unrepeatable as concerns their outcomes, as against the statistics of multiply repeated experiments, which are repeatable. It would be difficult, if not impossible, to pursue science without being able to repeat at least the statistical data our experiments provide. The absence of causality is an interpretative inference, automatic as it may be in the interpretations in the spirit of Copenhagen, each of which is, again, an idealization, albeit a nonrealist one.

It should be kept in mind that there are also causal or realist (or both) interpretations of quantum mechanics, or alternative quantum theories, such as Bohmian mechanics, or theories defined by an assumption of a deeper underlying (pre-quantum) causal dynamics, which makes quantum mechanics an "emergent" surface-level theory. Khrennikov's theory, discussed in Section 5, is a theory of this type, in which, however, not only quantum mechanics but also Bohr's view of measurement and, especially, his concept of complementarity (explained below) apply to quantum phenomena.[8]

One should also note that "the Copenhagen interpretation" is sometimes associated with the following view (e.g., [19]). The independent state of a quantum system is represented by the wave function that is associated with it. This wave function exhausts what could be known in advance about any possible state of the system. While the system is isolated from other systems (specifically measuring instruments), it evolves smoothly in time and its independent evolution is described, in realist and causal manner, by Schrödinger's equation, but it is unobservable. The lack of determinism manifested in quantum phenomena is due to the disturbance introduced by observation. Bohr briefly entertained this type of idea in the so-called Como lecture of 1927, although he was ambivalent about it even then [2, v. 1, pp. 54-55; 5, pp. 191-201]. The idea appears to originate with Dirac, who introduced it sometime in 1926 while at Bohr's Institute in Copenhagen and who appears to have influenced Bohr in this regard [20]. However, Bohr quickly came to realize the difficulties of sustaining it in quantum theory. Already his next publication on the subject abandons the idea, thus giving the Como argument barely a yearlong life span [2, v. 1, 92-101]. The view has, however, been and remains persistent in foundational arguments concerning quantum theory. This is not surprising. Apart from its general appeal on the account of realism and causality, it is found in Heisenberg's *The Principles of the Quantum Theory* [1], Dirac's *The Physical Principles of the Quantum Theory* [21], and especially J. von Neumann's *Mathematical Foundations of Quantum Mechanics* [22], all published around the same time (1930–1932). These were the most important early books on quantum mechanics, and they had and continue to have a strong impact. Although we will, in the discussion to follow, not be concerned with this form of the "Copenhagen" view and will restrict ourselves to Bohr's interpretation after 1927, which does not conform to this view and is indeed in conflict with it, we thought it was important to register it here and to comment on its persistence.[9]

---

[8] Among others who have pursued this line of inquiry are A. M. Cetto and L. de la Pena, M. Kupczynski, and T. Nieuwenhuizen (see [18]). Bohr came to see quantum field theory as extending quantum mechanics along a nonrealist gradient.

[9] For an extensive analysis of this view and problems found in it, see [5, pp. 191-219].

By "randomness" or "chance" we refer to a manifestation of the unpredictable. (Randomness and chance are not the same, but the difference between them is not germane for our argument.) It may or may not be possible to estimate whether a random event would occur, or even to anticipate it as an event. A random event may or may not result from some underlying causal dynamics unavailable to, or assumable by, us. Thus, in classical statistical physics, randomness and the resulting recourse to probability are due to insufficient information concerning systems that are at bottom causal but whose mechanical complexity prevents us from accessing their causal behavior and making deterministic predictions concerning this behavior. The situation is different in quantum physics, given the difficulties of sustaining arguments for the causality of the independent behavior of quantum objects or systems of quantum objects, even primitive (unsubdividable) individual quantum objects, such as those associated with elementary particles.[10] If an interpretation is nonrealist, the absence of causality is, again, automatic, and the recourse to probability is irreducible in principle rather than only in practice.

Probability and statistics deal with providing estimates of the occurrences of certain individual or collective events, in physics, or science in general, in accordance with mathematical probability theories. The terms "probabilistic" and "statistical" are generally used differently. "Probabilistic" refers to our estimates of the probabilities of either individual or collective events, such as of a coin toss or (assuming this claim is possible) of finding a quantum object in a given region of space. "Statistical" refers to our estimates concerning the outcomes of identical or similar experiments or to the average behavior of identical objects (or objects treated as identical).[11]

A given definition of probability may already reflect this difference, as is the case in the Bayesian vs. the frequentist definition of probability. The Bayesian understanding of probability defines it as a degree of belief concerning the occurrence of possible individual events on the basis of the relevant information we possess and hence, generally, is subjective. The frequentist understanding, sometimes also referred to as frequentist *statistics*, defines probability *statistically*, in terms of sample data by the emphasis on the frequency or proportion of these data, which is often seen as more objective, although this can be debated. The Bayesian approach allows one to make estimates even concerning individual and especially unique events, say, betting on the outcome of a basketball game or, as in Pascal's wager, on the existence of God and the salvation of the soul, rather than on frequently repeated events, such as repeated coin tosses. In the latter case, our estimations are defined by previous experience of the same or closely similar events; technically, no two coin tosses are ever quite the same, the point used by Bayesian theorists (there are differences between their views) against frequentist approaches to probability, which, again, reflects a more objectivist view [e.g., 24, pp. 317-320].

In quantum physics, where exact predictions appear to be, in general, impossible in principle even in dealing with individual objects and events, one considers identical quantum objects such as electrons or photons (not their identical preparation, which cannot be assured!), and the identically prepared measuring instruments as the initial condition of repeated experiments. The identical preparation of the instruments can be controlled (because their observable parts can be described classically), while that of quantum objects themselves, again, cannot be. As a result, the outcomes of quantum experiments, even if identically prepared in terms of the states of the measuring instruments involved, will, in general, be different, which fact could, as will be discussed below, be interpreted on either frequentist or Bayesian lines.

---

[10] The concept requires clarifications, which we put aside here. See [23], for a helpful discussion of this concept and the concept of quantum object in general from a realist perspective.

[11] The standard use of the term "quantum statistics" refers to the behavior of large multiplicities of identical quantum objects, such as electrons and photons, which behave differently, in accordance with, respectively, the Fermi-Dirac and the Bose-Einstein statistics.

This brief summary sidesteps some of the deeper aspects of probability, but it suffices for our purposes. We shall comment on some of these aspects later, especially in Sections 4 and 5.[12] We conclude here by noting that while randomness or chance introduces an element of chaos into order and reveals the character of world, or of our interactions with the world, probability introduces an element of order into situations defined by the role of randomness, and allows us to handle such situations better. Probability or statistics is about the interplay of randomness and order. This aspect of probability takes on a special, even unique, significance in quantum physics because of the presence of statistically ordered correlations (not found in classical physics) between certain data, such as those of the EPR-type experiments. These correlations are correctly predicted by the formalism of quantum mechanics and rules, such as Born's rule or various forms of the projection postulates, which are added to the formalism, rather than are inherent in it. This does not mean that it is the only formalism that can predict these correlations. Bohmian mechanics, the predictions of which coincide with those of standard quantum mechanics, predicts them as well, but at the expense of nonlocality, which standard quantum mechanics appear to avoid, although the issue is under debate.

Finally, by locality, we refer primarily to the compatibility with (special) relativity, which disallows physical connections or causal influences between events that propagate faster than the speed of light in a vacuum, $c$. Nonlocal theories, such as Bohmian mechanics (in all of its versions), allow for and even entail such connections. Bohmian mechanics allow for instantaneous connections of this type, even though it may not be possible to actually trace or enact these connections by human means.[13] Nonlocality in this sense is usually, albeit not always, seen as undesirable. Standard quantum mechanics appear to avoid it, although the question of the locality (in the sense of compatibility with relativity) of quantum mechanics or quantum phenomena is a matter of great subtlety and much controversy, especially, as noted earlier, in the wake of the Bell and Kochen-Specker theorems and related findings. There are alternative conceptions of nonlocality, sometimes linked to other terms and concepts, such as "separability" or "no signaling," some of which apply in standard quantum mechanics as well, even if the latter is interpreted in the spirit of Copenhagen. In particular, quantum mechanics may be seen as nonlocal insofar as it allows, specifically in the EPR-type situations, for *predictions* at an, in principle, arbitrary distance on the basis of measurements performed at a given location. However, as Bohr argued in his reply to the EPR [4], it is possible to interpret quantum mechanics so as to preserve locality, and thus to avoid what Einstein famously called in this connection "spooky action at a distance." In this view, while one could speak of spooky *predictions* at a distance, insofar as there is no physical explanation of the quantum-level reality responsible for these predictions, there is no spooky *action* at a distance [5, pp. 16, 271].

### 3. The Bohr-Einstein Debate and the Nature of Quantum Reality

---

[12] See, [25, 26] and references there. On the Bayesian philosophy of probability, in two different versions of it, see [24, 27].
[13] See Ref. [28] for an exposition of the last version of the theory developed by Bohm, in collaboration with B. Hiley.

Given the significance of the Bohr-Einstein debate for the subsequent discussions and debates concerning quantum foundations, we devote this section to outlining their respective positions and the essential issues at stake in their confrontation. While Einstein's position remained essentially stable throughout his life, Bohr's views underwent considerable revisions under the impact of his debate with Einstein and the development of quantum theory itself. Given our aims and scope, we shall only discuss the ultimate version of Bohr's view, stabilized sometime in the 1940s, by then especially shaped by his exchanges with Einstein concerning the EPR-type thought experiments (Einstein offered several versions of them).[14]

In characterizing Einstein's requirements for a proper, "complete," physical theory W. Pauli notes the precedence of realism over causality in Einstein's thinking [31, p. 13]. Intriguingly, Pauli does not comment on the significance of locality for Einstein, whose concept of realism was that of local realism. Einstein would prefer causality as well, in accordance with the ontological architecture of classical physics and relativity, in which his philosophical position was grounded. Pauli, however, is right to stress the precedence of realism over causality for Einstein. Pauli cites one of Einstein's many statements on the subject: "There is such a thing as a real state of a physical system, which exists objectively, independently of any observation or measurement, *and can in principle be described by the methods of expression of physics*" [32, p. 7; cited in 31, p. 131; emphasis added; also 3, pp. 138-39). The last part of Einstein's statement is important. It reflects his understanding of the idealized nature of our physical theories, which distinguishes his realism from naïve realism. Einstein's position is expressed more formally in the EPR paper in terms of a *necessary* criterion of completeness: for a theory to be considered complete (Einstein-complete) *"every element of the physical reality must have a counterpart in the physical theory"* [3, p. 138]. With this criterion in hand, EPR then formulated their famous *sufficient* criterion of reality: *"If, without in any way disturbing a system, we can predict with certainty (i.e., with probability equal to unity) the value of a physical quantity, then there exists an element of physical reality corresponding to this physical quantity"* [3, p. 138]. Although not always expressly stated, both criteria are retained in Einstein's subsequent arguments, where the focus of his analysis shifts from the completeness of quantum mechanics to its locality. The completeness, the Einstein-completeness, of quantum mechanics is still at stake in these arguments, just as locality was at stake in the EPR paper, because quantum mechanics is argued by Einstein to be either incomplete or, if complete, then nonlocal, the alternative considered in EPR's article as well [3, p. 141]. While his overall view concerning the completeness (the Einstein-completeness) of a physical theory remained in place throughout, Einstein's argumentation concerning the EPR-type experiments was qualified in his subsequent communications by statistical considerations to be discussed in the next section.

A crucial point is that, in this view, a proper quantum theory should be able to assign to a quantum system, say, an electron, properties, "elements of reality," that are independent of measurement and that would enable the theory to predict the electron's behavior on the basis of this assignment, ideally, in exact rather than probabilistic terms. This would imply causality. By the same token, the uncertainty relations (seen by Einstein as an artifact of quantum mechanics, reflecting its incompleteness, rather than a law of nature, as they were by Bohr) would not apply at the ultimate level. Hence, just as in classical physics or relativity, the theory would be ideally descriptive first and would make its predictions, again, ideally exact, on the basis of the description, "expression," it provides. Such a theory should, again, also be local, according to Einstein. The idea of "expression" in this sense is important for Einstein, because such an expression can only be achieved by means of conceptual construction, "the free choice of concepts," rather than by means of observable facts themselves, which Einstein sees as the empiricist "philosophical prejudice," found, for example, in E. Mach's philosophy [33, p. 47]. Einstein was not a naïve realist, given the role he assigns to a free choice of (mathematical) concepts in creating our theories and approaching physical reality, although he appears to have had more confidence than other realists, such as Schrödinger, in the power of our concepts to capture this reality with more exactitude [13, pp. 152-53]. Indeed, from the time of his creation of general relativity on, Einstein was progressively putting more and more faith, in his own work and in general, in the mathematical conceptual architecture of fundamental physical theories as the way of approaching, in a realist way, the ultimate nature of reality.

---

[14] For an extensive discussion of Bohr's interpretation, especially his ultimate interpretation, see [29]. For an insightful and helpful analysis of Bohr's earlier argumentation, see [30].

It is, Einstein argues, this kind of realist expression that quantum mechanics fails to deliver, which makes it incomplete, Einstein-incomplete, unless it is nonlocal. In other words, quantum mechanics fails in local realism. Einstein is factually correct, insofar as quantum mechanics, at least in Bohr's and related interpretations in the spirit Copenhagen, does not provide a physical "expression," a description, of individual quantum objects and processes, and it is the absence of this description that, as Bohr argued in his reply to EPR, allows one to maintain locality. In other words, while quantum mechanics may not be Einstein-complete, it is Bohr-complete. It is as complete as nature allows a theory of quantum phenomena to be, as things stand now, and, Bohr argues, it may also be interpreted as local [2, v. 2, p. 57; 4, pp. 701-702].

Bohr's interpretation is grounded in his argument concerning the irreducible role of measuring instruments in quantum physics, which eventually led him to his concept of phenomenon, defined by this role, introduced in the Warsaw Lecture of 1938 [34], under the impact of his exchanges with Einstein concerning the EPR experiment, and he based his ultimate interpretation in this concept. According to Bohr:

> I advocated the application of the word phenomenon exclusively to refer to the observations obtained under specified circumstances, including an account of the whole experimental arrangement. In such terminology, the observational problem is free of any special intricacy since, in actual experiments, all observations are expressed by unambiguous statements referring, for instance, to the registration of the point at which an electron arrives at a photographic plate. Moreover, speaking in such a way is just suited to emphasize that the appropriate physical interpretation of the symbolic quantum-mechanical formalism amounts only to predictions, of determinate or statistical character, pertaining to individual phenomena appearing under conditions defined by classical physical concepts [describing the relevant observable parts of measuring instruments]. [2, v. 2, p. 64]

Bohr's appeal to classical concepts is often misunderstood. While the subject requires a separate discussion, its key points are worth stating here. First, although indispensable, classical concepts are never sufficient for a proper account of quantum phenomena. Secondly, the interaction between quantum objects and measuring instruments is quantum (otherwise no measurement could take place), hence, is not amenable to a description in terms of classical or any other concepts, anymore than any quantum process is. In other words, measuring instruments have both the classical strata, which we observe, and the quantum strata, through which they interact with quantum objects. This interaction is quantum but is, in Bohr's terms, irreversibly amplified to the classical level of observation, a conception related, although not quite identical, to that of decoherence [2, v. 2, p. 73]. The preparation of measuring instruments is classical and thus controllable, which enables us to repeat it, and this preparation itself need not involve *h*. The outcomes of measurements, under the impact of quantum objects, while observed by classical means, will involve *h*. By contrast, measurements dealing with classical objects, never involve *h*.

Bohr sees quantum-mechanical formalism as "symbolic" because, while the theory uses mathematical symbols *analogous* to those used in classical mechanics, it does not, in his interpretation, refer these symbols to and, hence, does not describe the actual behavior of, quantum objects. Part of Bohr's concept of phenomenon is that this concept, in principle, precludes any description of quantum objects and their behavior, which also implies the irreducible difference between quantum phenomena and quantum objects [4, p. 700].[15] Nevertheless, it is this behavior that is responsible for the emergence of these phenomena.

A phenomenon in Bohr's sense, then, would refer to what is observed in a single experiment, and only in a single experiment, a crucial point, on which we shall further comment presently. Physical quantities obtained in quantum measurements, such as those defining the physical behavior of certain (classically described) parts of measuring instruments, are *effects* of the interactions between quantum objects and measuring instrument, a form of efficacy without causality. But these properties are no longer assumed to correspond, even in principle, to any properties pertaining to quantum objects, even any single such property, rather than only certain joint properties, in accordance with the uncertainty relations. Accordingly, the quantum-mechanical formalism only enables one to predict, in general probabilistically or statistically, what will be observed in measuring instruments under the impact of quantum objects.

It is crucial that the concept of phenomenon is defined by "the observations [already] *obtained* under specified circumstances" and hence only to already *registered* phenomena, rather than to what could be predicted. For one thing, such predictions are, *in general*, probabilistic or statistical and, hence, what will happen can never be assured. (As Bohr noted in the passage in question, in some experiments, such those of the EPR type, such predictions could, for all practical purposes, be seen as determinate, but, as will be discussed below, still not completely.) By the same token, the concept of phenomenon entails a rigorous specification of each arrangement, determined by the type of measurement we want to make. This specification also reflects the irreducibly individual, indeed unique and unrepeatable, character of each phenomenon, sometimes also referred to by Bohr as "atomicity." This concept is essentially equivalent to that of phenomena and should not be confused with the classical, Democritean, concept of atomicity, which refers to the indivisible ultimate constituents of matter itself. Bohr's "atomicity" only refers to phenomena associated with single experiments. The constitutive role of these conditions can never be eliminated in considering quantum measurement or predictions, including in the EPR-type experiments, where these predictions concern objects that are physically unaffected by the measurements enabling them [4].

---

[15] Technically, as Kant realized, objects and phenomena are also different in classical physics. There, however, this difference could be disregarded. More complex classical systems, such as those considered in classical statistical physics, introduces further complexities, such situations are still fundamentally different from those of quantum physics. This is because the elemental individual constituents of such systems could be treated by descriptive causal models of classical mechanics, which is impossible in quantum theory, at least in Bohr's and related interpretations.

Thus, if seen independently of the quantum mechanical context of its appearance, each mark on the screen in the double-slit experiment would be perceived as the same entity. Such a mark would appear the same regardless of the difference in the physical conditions and, hence, the outcome or rather the outcome*s*, "interference" or "no interference," of the experiment, are defined collectively by the two complementary setups of the experiment. The first is that with both slits open and no counters, which would allow us to know through which slit each quantum object passed, and the second is that when such a knowledge is possible in one way or another (for example, by using counters), even in principle rather than only actually. According to Bohr's understanding, however, each mark is part of a different individual phenomenon depending on these conditions, which are mutually exclusive in the case of complementary phenomena and are defined by each phenomenon uniquely in any circumstances. While, thus, a given single event does not allow one to establish in which setting it had occurred, the statistical distribution of the traces on the screen will always be different in these two setups. That the difference between two sets of outcomes is manifested only collectively over a large number of trials is important in interpreting the nature of probability assignments or, as in the statistical Copenhagen interpretation, the impossibility thereof in predicting the outcome of each individual experiment.

That we always have a free choice as concerns what kind of experiment we want to perform is in accordance with the very idea of experiment, which, as Bohr notes, also defines classical physics or science in general [4, p. 699]. Contrary to the case of classical physics, however, implementing our decision concerning what we want to do will allow us to make only a certain type of prediction (for example, that concerning a future position measurement) and will unavoidably exclude the possibility of certain other, *complementary*, types of prediction (in this case, that concerning a future momentum measurement). It is in this way that we actively shape what will happen, define the course of reality, as just noted.

Bohr's *concept* of complementarity reflects this mutual exclusivity of certain situations of measurement or phenomena, but is not restricted to it. Complementarity is defined by:

    *(a)* a mutual exclusivity of certain phenomena, entities, or conceptions; and yet
    *(b)* the possibility of applying each one of them separately at any given point; and
    *(c)* the necessity of using all of them at different moments for a comprehensive account of the totality of phenomena that we must consider.

Parts *(b)* and *(c)* of this definition are just as important as part *(a)*, and to miss them, as is often done, is to miss much of the import of Bohr's concept. Moreover, as just explained, in Bohr's interpretation (at least in the ultimate version of it), one only deals with complementary phenomena manifest in measuring instruments and their classical properties under the impact of quantum objects. One never deals with complementary properties of quantum objects or their independent behavior, because, in Bohr's view, no attribution of such properties, single or joint, to quantum objects is possible in the first place. Indeed, no complementary arrangements or phenomena can ever be associated with a single quantum object. One always needs two quantum objects in order to enact, in two separate experiments, two complementary arrangements, say, those associated, respectively, with the position or the momentum measurement, with the measured quantity itself physically pertaining strictly to the measuring instrument involved. The uncertainty relations, too, apply to the corresponding variables physically pertaining to measuring instruments and not to quantum objects. Furthermore, as correlative to complementarity, the uncertainty relations mean that one cannot even define, rather than only measure, both variables simultaneously. The recourse to probability, to which the uncertainty relations are correlative, is the cost of our active role in *defining* physical events in quantum physics, rather than merely tracking them, as in classical physics.[16]

Importantly, our freedom of choosing the experimental setup only allows us to select and control the initial setting up of a given experiment but not its outcome, which, again, can only be probabilistically estimated. This fact reflects the "objectivity" of the situation, defined by the verifiability and, thus, the possibility of the unambiguous communication of the data involved in our experiments (both that of their setups and their outcomes), and hence, the objective character of quantum mechanics in this interpretation. According to Bohr:

> A most conspicuous characteristic of atomic physics is the novel relationship between phenomena observed under experimental conditions demanding different elementary concepts for their description. Indeed, however contrasting such experiences might appear when attempting to picture a course of atomic processes on classical lines, they have to be considered as complementary in the sense that they represent equally essential knowledge about atomic systems and together exhaust this knowledge. The notion of complementarity does in no way involve a departure from our position as *detached observers* of nature, but must be regarded as the logical expression of our situation as regards objective description in this field of experience. The recognition that the interaction between the measuring tools and the physical systems under investigation constitutes an integral part of quantum phenomena has … forced us … to pay proper attention to the conditions of observation. [2, v. 2, p. 74; also p. 73; emphasis added]

Thus the observers in quantum physics are as detached from *measuring instruments* as they are in classical physics from *classical objects*, thus ensuring the objectivity of Bohr's scheme; the data in question or our predictions based on this data are the same for and hence independent of any particular observer. On the other hand, the measuring instruments used in quantum measurement can, in an act of observation or measurement, never be "detached" from quantum objects because the latter cannot be "extracted from" the *closed* observed phenomena (in Bohr's sense) containing them [2, v. 2, p. 73]. Phenomena *cannot be opened* so as to reach quantum objects by disregarding the role of measuring instruments in the way it is possible in classical physics or relativity, and thus are in conflict with Einstein's ideal of objectivity or completeness, which require a theory to describe the properties of the systems considered independently of measurement.

Hence, although quantum objects do *exist independently* of us and of our measuring instruments, they *cannot be observed or described independently*. Nobody has ever seen, at least not thus far, a moving electron or photon as such, but only traces of this "movement" (assuming even this concept applied), traces that, in view of the uncertainty relations, do not allow us to reconstitute this movement itself in the way it is possible in classical physics or relativity. In Bohr's interpretation, quantum phenomena, again, preclude any description, if not, as in the present view, a conception, of quantum objects themselves and their behavior, which behavior is, nevertheless, responsible for the emergence of these phenomena. As he said: "in quantum mechanics [in this interpretation] we are not dealing with an arbitrary renunciation of a more detailed analysis of atomic phenomena, but with a recognition that such an analysis is *in principle* excluded" [2, v. 2, p. 62]. Bohr noted that, while, Einstein's attitude "may seem well balanced in itself," it "implies a rejection," on philosophical grounds, of Bohr's argumentation leading to this conclusion [2, v. 2, p. 62].

---

[16] In part in view of the considerations given here, wave-particle complementarity, with which the concept of complementarity is associated most commonly, did not play a significant, if any, role in Bohr's thinking. Indeed, Bohr does not appear to have ever spoken of this complementarity. His solution to the dilemma of whether quantum objects are particles or waves—or his "escape" from the paradoxical necessity of seeing them as both—is that they are neither. Instead, either feature is seen by Bohr as an *effect* or set of *effects*, particle-*like* (which may be individual or collective) or wave-*like* (which are always collective), of the interactions between quantum objects and measuring instruments, in which these effects are observed.

It is sometimes argued that Bohr's interpretation implies that quantum mechanics or even physics in general should not be concerned with the ultimate character of nature, or reality, but only with what can be known about nature. Bohr did says on one occasion that in "our description of nature the purpose is not to disclose the real essence of the phenomena but only to track down, so far as it is possible, relations between the manifold aspects of our experience" [2, v. 1, 18]. It is worth keeping in mind, however, that this statement, actually uncommon, if not unique, in Bohr, reflects Bohr's 1929 and pre-EPR view, rather than his ultimate view. However, even if one accepts this type of view, it is, one might argue, not so much that Bohr wants to arbitrarily, merely as a matter of his philosophical position, renounce "disclosing the real essence of the phenomena," which could, after all, be part of our experience. Instead, one might argue that Bohr came to think, especially following his exchange with EPR, that, insofar as quantum mechanics and higher-level quantum theories are correct, they suggest that a quantum-level descriptive idealized model of quantum-level reality, *may be* "*in principle* excluded," [2, v. 2, p. 62]. This, again, need not mean that such a model is bound to remain excluded (hence, our emphasis on "may be"). As far as it is excluded, however, Bohr's position reflects a much stronger view than that urging one merely to renounce "disclosing the real essence of the phenomena," because, if such is the case, even if one wanted to have such a model, one might not be able to develop it. His full sentence makes this view even more pronounced: "in quantum mechanics, we are not dealing with an arbitrary renunciation of a more detailed analysis [reaching quantum objects and their behavior] of atomic phenomena but with a recognition that such an analysis is *in principle* excluded" [2, v. 2, p. 62]. By the same token, it is no longer possible "to disclose the real essence of the [quantum] phenomena," although it is possible "to track down, so far as it is possible, relations between the manifold aspects of our experience," and such a tracking would not be possible otherwise. In sum, Bohr was concerned with reality as much as Einstein was. The difference is that, unlike Einstein, Bohr came to accept the possibility of reality without realism.

## 4. Quantum Probability and the Statistical Copenhagen Interpretation

Bohr's epistemology, just outlined, entails a type of understanding of quantum randomness and quantum correlations, and of probability or statistics in quantum physics that is different from that found in classical physics, including classical statistical physics. This general understanding, however, allows for different interpretations, two of which will be considered in this section: Bohr's interpretation in its ultimate version, introduced in the preceding section, the statistical Copenhagen interpretation. We shall also, briefly and by way of comparison, comment on the Bayesian view of quantum probability, which could in turn allow for different interpretations of quantum mechanics, possibly related to different philosophies of probability, noted above.

The general understanding in question in this section is defined by the absence of realism and, as a consequence, the applicability of the idea of causality to individual quantum processes, as discussed in the preceding section in the context of Bohr's interpretation. This understanding, thus, fundamentally departs from the way we understand classical physical systems and processes in considering which the recourse to probability or statistics becomes necessary, as in classical statistical physics or chaos and complexity theories. In these cases the behavior of the physical systems considered is assumed to be causal. However, the mechanical complexity of these systems makes the recourse to probability or statistics unavoidable in predicting their behavior. A coin toss is, arguably, the most common example of this situation, unless the quantum aspects of the constitution of the coin are considered a factor (which is, generally, not the case).

By contrast, in dealing (in accordance with the understanding in question) with quantum objects and processes, we confront the absence of causality even in the case of primitive (indecomposable) individual objects ("elementary particles") and processes, which makes the probabilistic character of our predictions concerning *the corresponding quantum phenomena* unavoidable. The reference to the corresponding quantum phenomena and our emphasis reflect the fact that, *in this view or the interpretations of quantum objects and processes it implies*, our predictions only concern the effects of quantum processes manifested in the measuring instruments involved. This is an interpretation of the situation (hence our emphasis), which, as stated from the outset, is, as the situation, strictly in accord with what is observed, because, in Bohr's words, "one and the same experimental arrangement may yield different recordings [of their outcomes]" [2, v. 2, p. 73]. As explained above, it is possible to speak of "one and the same experimental arrangement," because, unlike the outcomes of experiments, we can control the measuring instruments involved, given that the observable parts of these instruments relevant for setting up our experiments can be described and (with qualifications offered earlier) behave classically. On the other hand, the state of each quantum object under investigation in each repeated experiment (say, at the time when an electron or photon is emitted from the source considered) will not, in general, be identical. Under these conditions, the probabilistic character of such predictions will, again, concern even elemental individual quantum events. For, in contrast to classical physics, in the case of quantum phenomena it does not appear possible (and in Bohr's and related interpretations is rigorously impossible) to subdivide these phenomena into entities of different kinds concerning which our predictions could be exact, even ideally or in principle. Any attempt to do so will require the use of an experimental setup that leads to a phenomenon or set of phenomena of the *epistemologically* same type (they could be different physically), concerning which we could again only make probabilistic predictions, or possibly only statistical predictions that concern multiplicities of identically prepared experiments.[17]

---

[17] Given the data obtained in quantum experiments, this would have to be the case even if one assumes the underlying causality of quantum processes. This fact is reflected in Bohmian theories, in which a given quantum object is assumed to possess both position and the momentum, as defined exactly, at any moment of time, thus allowing one for realism and causality. However, these theories retain the uncertainty relations and, correlatively, reproduce the statistical predictions of quantum mechanics, because a given measurement always disturbs, *actually disturbs*, the object and displaces the value of one of these properties. By contrast, this type of concept of disturbing quantum objects and processes by observation (a concept that allows one to give a classical-like independent architecture to quantum objects and processes

Bohr presents his view of quantum probability in terms of the concepts of phenomenon and atomicity (which are, as noted, more or less equivalent) in the Warsaw lecture of 1938, "The Causality Problem in Atomic Physics," which introduces both concepts. He says:

> The unrestricted applicability of the causal mode of description to physical phenomena has hardly been seriously questioned until Planck's discovery of the quantum of action, which disclosed a novel feature of atomicity in the laws of nature supplementing in such unsuspected manner the old doctrine of the limited divisibility of matter. Before this discovery statistical methods were of course extensively used in atomic theory but merely as a practical means of dealing with the complicated mechanical problems met with in the attempt at tracing the ordinary properties of matter back to the behaviour of assemblies of immense numbers of atoms. It is true that the very formulation of the laws of thermodynamics involves an essential renunciation of the complete mechanical description of such assemblies and thereby exhibits a certain formal resemblance with typical problems of quantum theory. So far there was, however, no question of any limitation in the possibility of carrying out in principle such a complete description; on the contrary, the ordinary ideas of mechanics and thermodynamics were found to have a large field of application also proper to atomic phenomena, and above all to offer an entirely sufficient basis for the experiments leading to the isolation of the electron and the measurement of its charge and mass. Due to the essentially statistical character of the thermodynamical problems which led to the discovery of the quantum of action, it was also not to begin with realized, that the insufficiency of the laws of classical mechanics and electrodynamics in dealing with atomic problems, disclosed by this discovery, implies a shortcoming of the causality ideal itself. [34, pp. 94–95]

As Bohr says elsewhere: "[I]t is most important to realize that the recourse to probability laws under such circumstances is essentially different in aim from the familiar application of statistical considerations as practical means of accounting for the properties of mechanical systems of great structural complexity. In fact, in quantum physics we are presented not with intricacies of this kind, but with the inability of the classical frame of concepts to comprise the peculiar feature of indivisibility, or 'individuality,' characterizing the elementary processes" [2, p. 34]. This inability is primarily due to the inaccessible or even (although Bohr, again, may not have claimed as much) inconceivable nature quantum processes. The lack of causality, which is, again, an automatic consequence, limits us to probabilistic or possibly only statistical estimates of the outcomes of all quantum experiments. Thus, in each case, the wave function provides, in Schrödinger's way of putting it (by this point, in 1935, disparagingly), probabilistic or statistical "expectation catalogues" concerning quantum experiments [13, p. 158]. Any such catalogue is reset with each new measurement, which renders previous history of measurement on the same object irrelevant as concerns our predictions from this point on [5, pp. 73-76]. The meaning of these catalogues may be subjected to a further interpretation, even if one follows the spirit of Copenhagen. It may be added that, if one thinks along these lines, the wave function does not have any special significance vis-à-vis other ways of providing quantum predictions. In quantum field theory, one no longer relies on wave functions (which are Hilbert space vectors) and uses only Hilbert-space operators. At the same time, however, the fact that probabilistic predictions of quantum mechanics are correct is enigmatic, insofar as it does not appear to have, and in Bohr's

---

when they are not disturbed by observation) is inapplicable in Bohr's interpretation or the statistical Copenhagen interpretation [2, v. 2, pp. 63-64].

view does not have, an underlying physical justification of the type found in classical physics when the latter must use probability and statistics. We appear to be lucky to be able to make these predictions. According to Pauli, who follows Bohr and even uses Bohr's locutions ("rational generalization" and "the finiteness of the quantum of action"), although Pauli's overall interpretation of quantum phenomena and quantum mechanics is somewhat different:

> As this indeterminacy [that reflected in the uncertainty relations] is an unavoidable element of every initial state of a system [a quantum object] that is at all possible according to the new [quantum-mechanical] law, the development of the system even can never be determined as was the case in classical mechanics. The theory predicts only the *statistics* of the results of an experiment, when it is repeated under a give condition. Like the ultimate fact without any cause, the *individual* outcome of a measurement is, however, in general not comprehended by laws. This must in general be the case, if quantum or wave mechanics is interpreted as a rational generalization of classical physics [mechanics], which take into account the finiteness of the quantum of action [Planck's constant, $h$]. The probabilities occurring in the new laws have then to be considered to be primary, which means not deducible from deterministic [causal] laws. As an example of these primary probabilities I mention here the fact that the time at which an individual atom will undergo a certain reaction stays undetermined even under conditions where the rate of occurrence of this reaction for a large collection of atoms is practically certain. [31, p. 32][18]

Pauli speaks of determinism rather than causality, but this does not affect the situation in question because his concept of determinism is in effect the same as the present concept of causality and, besides, as explained in Section 2, in the case of classical mechanics both notions coincide. Most important here is Pauli's main claim: "The theory predicts only the *statistics* of the results of an experiment, when it is repeated under a give condition. Like the ultimate fact without any cause, the *individual* outcome of a measurement is, however, in general not comprehended by laws." Given that Pauli does not specify otherwise, this appears to include probabilistic or, in this view, only *statistical* laws of quantum mechanics, which, by definition, only apply to statistical multiplicities of repeated quantum events. Indeed, he corroborates this reading elsewhere in the context of complementarity: "In the general case of the quantum-mechanical state of a material particle, neither the position nor the momentum is predictable with certainty; in consequence the state can be described only by *statistical* statements about the distribution of values of the results of possible measurements of position or momentum of this state. Formally these statements are embraced symbolically in a wave function, consisting of a real and an imaginary part" (31, p. 99; Pauli's emphasis).

This point has an important implication, which is not stated by Pauli in this form but which leads to the statistical Copenhagen interpretation. For it follows that, in this view, the primitive individual quantum processes and events are not only beyond description, that of or derived from quantum-mechanical formalism included, or possibly even conception, but are also beyond predictions, even probabilistic predictions (exact predictions, even ideal ones, are, again, excluded automatically in the absence of such a description). In other words, the outcome of an individual quantum experiment, a future individual quantum event, cannot, *in general*, be assigned probability: it is completely random. Only the statistics of multiply repeated (identically prepared) experiments can be predicted, which gives the corresponding meaning to the expectation catalogues provided by the formalism, say, by the wave function. We shall explain the emphasis on "in general" presently, merely noting for the moment that in some experiments, it is possible to assign a probability to individual quantum events for all practical purposes, but not in full rigor. It might appear or be argued that one could speak more rigorously (rather than only loosely) of the probability of a (future) individual quantum event, similarly to the way one does in the case of a coin toss. It might also appear that, just as in the case of a coin toss, one could, *by using the quantum-mechanical formalism, say, the wave function*, assign probabilities for individual events, defined, along Bayesian lines, in terms of "a degree of belief," rather than in terms of frequencies or statistics. Both types of claims would appear to be implied by the standard Bayesian views of probability, found, for example, in [24] and [27]. (Such views may deviate from each other in other respects.) It is not our aim to unconditionally deny these claims or argue against Bayesian approaches to quantum phenomena and quantum mechanics, which need not be limited by these claims alone.[19] We argue instead that these particular claims pose difficulties given the data observed in quantum experiments thus far, and that in any event the statistical view of quantum phenomena and quantum mechanics (advocated by both interpretations proposed here, different as they are in other respects, specifically as concerns realism) is consistent with the character of these data.

---

[18] Among other major figures who adopted this position were J. Schwinger [35, pp. 14-15] and, earlier, again, with a negative attitude, Schrödinger [13, p. 154].

The statistical Copenhagen interpretation may be seen (a matter of perspective, however) as more radical epistemologically than the Bayesian view, even if a nonrealist one, as defined by the claims just stated, insofar as in this interpretation not even the probabilistic "knowledge" concerning, or even the application of the concept of probability to, individual quantum events is possible. First of all, at least in most quantum situations, any verification of such individual estimates would still involve multiple events, and these individual estimates will rely on that data, reflecting these repeatable statistics. A Bayesian might contest this point in the case of quantum phenomena similarly to the way it might be and has been contested by Bayesians in the case of our estimates of the probability of a coin toss [24, pp. 317-20]. More crucial is that an individual quantum event may be, and in the statistical Copenhagen interpretation is, beyond assigning it a probability at all. Consider the double-slit experiment. It is true that in the case of the interference setup (both slits open and there are no counters allowing us to detect through which slit each particle in question passes), we observe the interference pattern (in the absence of actual waves, "correlational pattern" might be a better term), which is defined by the zones of permitted or "forbidden" impact. This pattern is strictly statistical in nature, manifesting itself only in a very large number of trials, around 70,000. The statistical nature of this pattern also reflects the fact that rigorously there is no zone of "forbidden" impact for any individual trial. Any given trial can leave its mark anywhere on the screen, as is clearly shown by the famous data of A. Tonomura's single electron build-up experiments [38]. Some trials (admittedly a statistically small number) can produce no impact at all.

---

[19] The so-called quantum Bayesianism or QBism exemplifies the complexities of Bayesian thinking in quantum theory and beyond [36, 37]. While it adopts a nonrealist view of quantum objects and processes, and *in this respect* is in accord with Bohr's interpretation and the statistical Copenhagen interpretation, it is different from both in other respects [37]. QBism is Bayesian and nonrealist, but not all Bayesian or all nonrealist positions are QBist. To properly address QBism, including in its Bayesian aspects, and to fairly assess its claims would require an extensive analysis that cannot be undertaken here.

In view of these considerations, it appears difficult to speak *rigorously* of assigning a probability to an individual trial. It is true that given the setups in certain experiments, such as the Stern-Gerlach experiment or some interferometry experiments it may be possible to speak, nearly with certainty, of an object having a 50 percent probability of taking one path or another in the corresponding arrangement. First, however, if so, this is true only in some experiments, which is why we use "in general" in speaking of the impossibility of such assignments. Secondly, even in these cases, there will be some trials, however small in number, in which no outcome will be registered, and unlike in classical cases, where similar statistical deviations may occur, it is not a matter of outside interferences, but something inherent in quantum experiments.

It would, then, also appear (without, again, offering a definitive claim) that these circumstances complicate, even if not exclude, the application, in quantum physics, of a Bayesian view, insofar as it refers estimates, bets on, the outcomes of individual events on the basis of the information one has.[20] A given quantum phenomenon or event would, in Bohr's definition, be seen in relation to the conditions defining this multiplicity, such as one or the other setups of the double-slit experiment, that gives rise to the interference patter or that in which this pattern will not appear. According to Pauli: "The mathematical inclusion, in quantum mechanics, of the *possibilities* of natural events has turned out to be a sufficiently wide framework to embrace the *irrational* actuality [i.e., beyond even the laws of quantum theory] of the single event as well" [31, p. 47]. In any event, a corresponding interpretation, such the statistical Copenhagen interpretation, appears to be consistent with the experimental data in question in quantum mechanics.

Finally, to exercise an even greater caution, it is also true that one could, even in quantum experiments (we, again, only addressing quantum physics here), assign probabilities on other bases than those of quantum mechanics, for example, by using other physical laws or even without any physical laws. We are also not saying that the Bayesian approach does not work in general in physics and beyond: there are many situations where it does, for example, when we need estimate the probabilities of certain human events, as in betting on the outcome of a basketball game. Even in quantum experiments, then, one could make any predictions one likes or must in view of one's Bayesian prior. However, the question is that of the effectiveness of our predictions in physics, of who will do, bet, better in physics, in predicting the outcomes of physical experiments. Consider, again, the double-slit experiment. One will "win" with quantum mechanics in hand against those who do not know it or whose theory is not as good, but one would win or will not consistently lose, only in many trials, which fact implicitly contains statistics. This is far from insignificant, and is a powerful reason to use quantum mechanics for predicting the outcomes of quantum experiments. Our point is only that quantum mechanics, generally, offers one no help as concerns predicting and hence betting on the outcome of a single experiment, because one cannot rigorously predict what happens, although, as noted earlier, this is in practice workable in some experiments—in practice, but not in full rigor. Accordingly, still speaking with more caution than definitiveness and by way of an interpretive choice, the statistical approach to quantum mechanics may be more rigorous than a Bayesian one. Pauli appears to have thought so as well. In fact, none of "the usual Copenhagen suspects" appears to have been Bayesian.

---

[20] Of course, in the Bayesian scheme of things, there may be some individual events to which one cannot assign probabilities, but, in general, one can and usually does.

Bohr's position concerning this alternative between assigning probabilities to individual quantum events on Bayesian lines and the strictly statistical nature of all quantum predictions (the statistical Copenhagen interpretation) does not appear to have been expressly stated in his works. Hence it could, in principle, be interpreted either way, and it was interpreted on more Bayesian lines previously by one of the present authors [5]. However, there does not appear to be an expressed statement to that effect in Bohr either. Indeed, Bohr, who is, again, careful in selecting his terms, clearly and even strictly prefers "statistical" to "probabilistic" in referring to quantum predictions throughout his oeuvre, which makes one inclined to see his view along the statistical, rather than Bayesian, lines, as against [5], thus revising the position taken there. Consider his comment, already cited above but especially relevant here, on Einstein's 1936 criticism of quantum mechanics:

> Einstein … argued that the quantum-mechanical description is to be considered merely as a means of accounting for the average behavior of a large number of atomic systems, and his attitude to the belief that it should offer an exhaustive description of the individual phenomena is expressed in the following words: "To believe this is logically possible without contradiction, but is so very contrary to my scientific instinct that I cannot forego the search for a more complete conception [that of the description of individual quantum processes, as an ideally exact description]." … Even if such an attitude might seem well balanced in itself, it nevertheless implies a rejection of the whole argument exposed in the preceding [essentially an argument explaining Bohr's interpretation], aiming to show that, in quantum mechanics, we are not dealing with an arbitrary renunciation of a more detailed analysis [reaching quantum objects and their behavior] of atomic phenomena but with a recognition that such an analysis is *in principle* excluded. [2, v. 2, pp. 61-62; 39, p. 349]

It might appear that Bohr implies here that quantum mechanics does, contrary to Einstein's assessment, provide an exhaustive "description" of the *individual* quantum phenomena, rather than as a means of accounting for the average behavior of a large number of atomic systems. More accurately, given his interpretation as considered here, Bohr might appear to suggest that quantum mechanics provides an exhaustive *predictive account* of individual quantum phenomena. However, he does not say that these predictions, which, it follows, would be unavoidably probabilistic, concern individual phenomena, rather than the statistics obtained in repeated identically prepared experiments, as, say, in the double-slit experiment. It is true that, as explained above, Bohr defines each phenomenon as individual. But he also defined a phenomenon only as an observed, registered phenomenon, and not as anything predicted. This allows one to interpret all quantum-mechanical predictions as, *in general*, statistical, rather than as probabilistic predictions pertaining to individual quantum experiments, consistently with this definition, "in general," because in some cases individual predictions are possible for all practical purposes, but never completely.[21]

---

[21] In certain situations, such as those of the EPR type, we can, for all practical purposes, predict certain quantities exactly, but this is never true in full rigor either, for the reasons just considered. There is always a non-zero probability that the object in question will not be found where it is expected to be found at the moment of time for which the prediction is made. Unlike the Bell-Bohm version of the EPR experiment for spin (at stake in Bell's and related theorems), the actual experiment proposed by EPR, dealing with continuous variables, cannot be physically realized, because the EPR-entangled quantum state is non-normalizable. This fact does not affect the fundamentals of the case, which can be considered in terms of the idealized experiment proposed

Accordingly, what Bohr says here is at the very least compatible with the statistical Copenhagen interpretation. His elaboration, then, could be read as follows. It is true that, at least in Bohr's or other interpretations in the spirit of Copenhagen, quantum mechanics does not provide a description of individual quantum processes (or quantum objects), and, if this is the criterion for the completeness of quantum theory, it is incomplete. The question, as noted from the outset, is whether nature or our interactions with nature allow us to do better. Bohr assumes that it might not, at least as things stand now, which is, again, not the same as to say that it never will. This is the meaning of his statement that "in quantum mechanics, we are not dealing with an arbitrary renunciation of a more detailed analysis [reaching quantum objects and their behavior] of atomic phenomena but with a recognition that such an analysis is *in principle* excluded." This is not inconsistent with the fact that quantum mechanics is "a means of accounting for the average behavior of a large number of atomic systems" in repeated experiments, a means of predictive accounts, providing the statistics of these repeated experiments. In other words, it is not only that individual (or for that matter any other) quantum objects and processes are *beyond description or even conception*, but, which is the main point at the moment, also that individual quantum experiments are *beyond prediction* as well, allowing only for the statistics of repeated experiments. Any analysis beyond that of phenomena in Bohr's sense (which are individual) and the statistics of repeated experiments (which are always multiple) is "*in principle* excluded," again, at least as things stand now.

This may be unacceptable to Einstein, but may, again, be as much as it is possible for us to have, which makes quantum mechanics complete, as complete as possible, at least as things stand now. Einstein might not have had in mind all the nuances just spelled out, but this is secondary, vis-à-vis the interpretation itself thus suggested. Einstein did note, however, that if quantum mechanics is the statistical theory of ensembles, then the paradox of nonlocality arising from his analysis of the EPR-type experiment would disappear, and quantum mechanics could be seen as local. As he said on several occasions, if one regards the wave function as relating to "many systems, to 'an ensemble of systems,' in the sense of statistical mechanics," then "the paradox" arising in view of EPR's argument is eliminated [39, p. 375; 32, p. 8; 40, pp. 205, 211].

Einstein's statistical alternative, however, still leaves quantum mechanics incomplete by his criteria, Einstein-incomplete, because of its inability to provide a properly exhaustive physical description of the behavior of individual quantum systems of the kind classical mechanics does, including for the individual constituents of the systems considered in classical statistical physics. Accordingly, this alternative would still be insufficient for him to accept quantum mechanics as the way of describing nature in its ultimate constitution. As he says, "[O]ne can safely accept the fact . . . that the description of the single system is incomplete, if one assumes that there is no corresponding complete law for the complete description of the single system which [law] determines its development in time" [40, p. 205]. In this case, "The statistical character of the ...

---

by EPR. There are experiments (e.g., those involving photon pairs produced in parametric down conversion) that statistically approximate the idealized entangled state constructed by EPR for continuous variables. These experiments are consistent with the present argument. They also reflect the fact that the EPR thought experiment is a manifestation of correlated events for identically prepared experiments with EPR pairs, which can in this regard be understood on the model of the Bell-Bohm version of the EPR experiment. In any event, there are quantum experiments, such as, paradigmatically, the double-slit experiment, in which the assignment of probabilities to the outcomes of individual events is difficult and even impossible to assume.

theory would ... follow necessarily from the incompleteness of the description of the [individual] systems in quantum mechanics" [33, p. 81]. In other words, in this view, quantum mechanics provides no account of individual quantum systems at all, which, again, corresponds to Bohr's view or that of the statistical Copenhagen interpretation. In this view, however, such a theory is complete insofar as it does all that nature or, again, our interactions with nature allows us to do. So both completeness (admittedly only the Bohr completeness, which leaves quantum mechanics Einstein-incomplete) and locality would be preserved, in either Bohr's interpretation, whether one interprets it on Bayesian or statistical lines, or in the statistical Copenhagen interpretation.

### 5. The PCSFT Model: Quantum Probabilities from Classical Random Fields Interacting with Threshold Detectors

We now present the statistical model, "the pre-quantum classical statistical field theory (PCSFT)," explained in detail in [41-43], to which we refer the reader for details. Here we only outline the key conceptual features of the model. In this model, quantum systems are considered as "symbolic" representations of classical random fields fluctuating at time and space scales that are essentially finer than quantum laboratory scales. In relation to this underlying level, this model is realist, analogously to classical statistical physics (which, however, deals with particle-like objects). We also keep in mind the complexities of realist models indicated earlier, which are present in this model as well but which we put aside. The key feature of the model as far as its potential relation to the experiment is concerned is that, by interacting with detectors of the threshold type, the classical fields considered by the model produce clicks in these detectors [42, 43, 45]. These clicks are interpreted as quantum events. They are context-dependent, just as such clicks are in standard quantum mechanics, in whatever interpretation, and, to begin with, in the observed quantum phenomena [41, 42].

They also correspond to the probabilistic or statistical predictions of quantum mechanics, and hence could, in PCSFT, be interpreted on lines of the statistical Copenhagen interpretation, as discussed earlier. We note, however, that the dependence of predictions of PCSFT on the threshold detection process involves numerous experimental parameters, such as the detection threshold, size of coincidence window, pulse duration, and so on [39, 40, 42]. These parameters are typically treated as technical experimental features, "experimental technicalities," and are ignored in other fundamental pre-quantum models. The presence of such technicalities as constitutive (rather than auxiliary, technical) elements is a strong feature of PCSFT, because it allows for the falsification of the model. By the same token, although the model's probabilistic predictions for frequencies of clicks of threshold type detectors coincide with those of quantum theory, the special type of dependence on the parameters just mentioned makes it possible to distinguish PCSFT from quantum mechanics at the level of such dependencies. Moreover, in many cases, limiting our understanding to the standard quantum formalism cannot provide definite recommendations concerning such technicalities. For example, it appears that in the Bell-type experiments the size of the coincidence window is not derived from quantum theory; experimenters select this size "by hand" to match better with quantum theory [46]. PCSFT typically predicts the range of technical parameters, for example, the coincidence windows, and hence can in principle be falsified on experimental grounds, without referring to quantum-mechanical formalism.

PCSFT carries with it a rather special interpretation of the wave function or, more generally, of the density operator as the covariance operator of the prequantum random field. This interpretation differs crucially from Schrödinger's initial conception of the wave function as explicitly representing a physical field or interpretations of quantum mechanics that maintain this conception.[22] On the other hand, it is in accord with Einstein's interpretation of the wave function as representing statistical features of an ensemble of identically prepared quantum systems, with treating them as symbolic representations of classical random fields. However, the interpretation of the quantum state as the covariance operator of the corresponding prequantum random field gives a specific physical content to Einstein's statistical interpretation, because it explicitly specify the way in which statistical features of the ensemble of identically prepared systems is encrypted in the quantum state. By the same token, it is also different from the statistical Copenhagen interpretation, which is equally statistical in that it refers to an ensemble of the identically prepared experiments (in terms of the measuring instruments involved), but in which the wave function or the density operator is a merely symbolic predictive tool devoid of any physical content. It should also be noted that, for composite quantum systems, the covariance operator of the corresponding random field contains an additional component corresponding to the random background field, i.e., this operator is not reduced to the quantum state. The role of the background random field is discussed in more detail below.

In PCSFT, the irradiance of a beam of light is only an indication of its average state. If we could magnify local states, we should see a little bit of chaos. At some points the amplitude of the waves is well below the average, and at other points we get arbitrarily high spikes. In other words, the field is "clumpy" at the microscopic level. Now, let us suppose that we have a point-like detector. When the field crosses the plane of detection, it might happen that the local amplitude is close to average or lower. No detection is possible. It can also happen that we have an amplitude spike followed by several small crests. Again, the signal does not accumulate above the threshold, and nothing happens. Yet, there is a real probability that an amplitude spike will continue over several cycles. In this case, sustained resonance above the threshold will result in a detection click. Consequently, the pattern of detection is produced by the low probability of transient "spikes" in a continuous field. Now, it is not true that we deal with single discrete entities at the moment and point of detection. The probability of a coincidence click, i.e., matching of two trains of spikes (at the micro-scale) at two detectors is nonzero, even theoretically. It decreases with increase of the threshold, but even for very high thresholds a random field can produce matching spikes. Thus, although, if one installs two detectors, one by each slit, the probability of double clicks can be made very small, they are ultimately irreducible [42]. This is one of the reasons why it is impossible to use the functional (as opposed to operator) representation of quantum observables. However, the main reason for the situation is quantum contextuality, corresponding to and indeed defining Bohr's concept of complementarity and his interpretation of quantum mechanics, as discussed earlier. A classical signal has no sharp position in space, that is, what is known as the "value definitiveness postulate" (all observables defined for a quantum-mechanical system have definite values at all times) is not valid for classical signals. "Signal's position" only has meaning in the context of the position measurement [43].

---

[22] Schrödinger retreated from the idea in the late 1920s in the wake of Born's probabilistic interpretation of it as a "probability catalogue," as Schrödinger himself called it, as explained earlier [13, p. 158]. However, he gradually returned to it following EPR's paper [3].

However, at least in this model, one cannot, in general, represent quantum compound systems in entangled states by considering random fields propagating in a vacuum, and, hence, say, to violate Bell's inequality in the way quantum mechanics or it appears (the case is as yet not entirely established experimentally or is at least under debate) the relevant observed data does. One has to consider a random background field that is present everywhere. We may note, by way of analogy (keeping in mind the differences between the present model and Schrödinger's conception), that this is not unlike the way in which Schrödinger thought, on his way to the discovery of his version of quantum mechanics as wave mechanics, of "the wave radiation that forms the basis of the universe." In Schrödinger's way of thinking, the concept of an elementary particle, mathematically considered by then, as it is now, a dimensional point-like entity, would be replaced with that of a particular effect of these wave-like vibrations. As he said: "This means nothing more than taking seriously the undulatory theory of the moving corpuscle proposed by de Broglie and Einstein, according to which the latter [i.e. the corpuscle] is nothing more than a kind of 'white crest' on the wave radiation forming the basis of the universe" [48, p. 95]. This became the program he pursued in his approach to quantum phenomena. At bottom, everything would be continuous, field-like. Accordingly, his mechanics was an undulatory, *wave*, rather than *quantum*, mechanics. It was analogous to classical wave physics but, as he stressed, not identical to the latter, in part, given the difference between his mathematical formalism and that of classical (wave) physics.

In the present model, by contrast, the deeper underlying field (absent in Schrödinger's approach) is classical. This might be either in accord with or in contrast to other approaches of, conceptually, the same kind, such as those of quantum-gravity proposals, depending on whether the underlying fields considered in such a proposal are classical, or classical-like, or not, for example, if it is more quantum-like. The status of string or brane theory in this regard is an intriguing question, which we shall, however, put aside here, although the scale of these theories is a relevant issue. We also leave aside the question of quantum field theory and high-energy fundamental physics from this perspective, which would require an extensive separate treatment.

One might add that PCSFT, as a theory of continuous classical fields, and hence the overall model under discussion is manifestly different from Bohmian theories, for one thing, because at the quantum limits they imply different theories, the standard quantum mechanics in the first case and Bohmian mechanics in the second. PCSFT matches well with the dream of A. Einstein and L. Infeld for a classical field model of physical reality [49]. As noted above, Einstein continued to pursue the project of developing such a model until literally the last days of his life. Of course, we cannot know whether Einstein and Infeld would accept PCSFT as one of possible realizations of their dream. They might have responded to it similarly to the way Einstein did to the Bohmian model (admittedly a particle-model, with a guiding—"pilot"—field, a concept originally suggested by L. de Broglie, and derived by him from Einstein's earlier ideas, abandoned by Einstein). Einstein found the Bohmian model interesting, but not in accordance with what he thought ideally required, in part because of nonlocality. By contrast, PCSFT is local in the same way as a theory of classical fields, such as, the classical electromagnetic field. There is no "action at a distance" that one finds in Bohmian theory. However, given that PCSFT reproduces the predictions of standard quantum mechanics, which is nonrelativistic, PCSFT is also a nonrelativistic theory. The presence of the underlying classical field may be interpreted as the presence of an underlying nonlocal structure contributing to correlations, which therefore should be considered as nonlocal ones, in the sense of spooky predictions, but not action, at a distance as discussed earlier. However, this underlying field is modeled in a fully classical manner, in the way the noisy background is modeled in classical radio-physics. This field's contributions to correlations are purely local, because they are based on field integration in small neighborhoods of detectors. Roughly speaking, the problem of locality in PCSFT has the same status as it has in classical radio-engineering.

One may call this underlying classical field the zero point field or vacuum fluctuations. This field has a random structure similar to that of random field-signals representing quantum systems. Hence, a threshold detector "eats" energy of combined spikes, signals combined with the background field. "Non-objectivity," in the sense of the context-dependence of measurement outcomes (they are "objective" in Bohr's sense, as explained above) of such observables on random fields is a consequence of the impossibility in general of assigning, say, polarization up or down prior to measurement. As a consequence of the presence of the random background field contributing irreducibly into threshold detection, the coincidence clicks appear irrespectively to our manipulations with random field-signals representing quantum systems. By the same token, this background field contributes to correlations and, in particular, its presence establishes a possibility for violating Bell's inequality, just as quantum mechanics does. Thus observable "quantum events," which are, again, clicks of detectors, cannot be assigned to pre-quantum physical systems themselves, which are classical fields [41, 42]. However, these systems have their own properties, not found in other quantum models, such as that of the standard quantum mechanics or that of Bohmian theories, features that might be, at least in principle, approachable and experimentally testable, which would change our conception and knowledge of physical reality beyond the quantum.

It is a safe bet that this conception and knowledge will change in one way or another. But in which way will they change? It may not be that different from asking which way an electron will go in the double-slit experiment. We can never be certain beforehand in either setup. But then, this is still thinking the way we do in quantum physics. We might need something else in physics and beyond.

**Acknowledgments.** The authors are grateful to Irina Basieva, Ceslav Brukner, G. Mauro D'Ariano, Henry Folse, Gregg Jaeger, Jan-Åke Larsson, and Anton Zeilinger, for exceptionally productive discussions, and to both anonymous readers of the article for their insightful comments and helpful suggestions.